\documentclass[aps,12pt,superscriptaddress,amsfonts,amssymb,amsmath]{revtex4}

\usepackage{epsfig}
\usepackage{makeidx}
\usepackage{graphicx}
\usepackage{epstopdf}
\usepackage{natbib}
\begin{document}

\title{The Bass diffusion model on finite Barabasi-Albert networks}

\author{M.L.\ Bertotti \footnote{Email address: marialetizia.bertotti@unibz.it}}
\author{G.\ Modanese \footnote{Email address: giovanni.modanese@unibz.it}}
\affiliation{Free University of Bolzano-Bozen \\ Faculty of Science and Technology \\ I-39100 Bolzano, Italy}

\linespread{0.9}

\begin{abstract}

\bigskip

Using a mean-field network formulation of the Bass innovation diffusion model and exact results by Fotouhi and Rabbat on the degree correlations of Barabasi-Albert networks, we compute the times of the diffusion peak and compare them with those on scale-free networks which have the same scale-free exponent but different assortativity properties. We compare our results with those obtained by Caldarelli et al.\ for the SIS epidemic model with the spectral method applied to adjacency matrices. It turns out that diffusion times on finite Barabasi-Albert networks are at a minimum. This may be due to a little-known property of these networks: although the value of the assortativity coefficient is close to zero, they look disassortative if one considers only a bounded range of degrees, including the smallest ones, and slightly assortative on the range of the higher degrees. We also find that if the trickle-down character of the diffusion process is enhanced by a larger initial stimulus on the hubs (via a inhomogeneous linear term in the Bass model), the relative difference between the diffusion times for BA networks and uncorrelated networks is even larger, reaching for instance the 34\% in a typical case on a network with $10^4$ nodes.

\end{abstract}

\maketitle

\section{Introduction}
\label{introduction}

The study of epidemic diffusion is one of the most important applications of network theory \cite{keeling2005networks}, the absence of an epidemic threshold on scale-free networks being perhaps the best known result \cite{pastor2001epidemic}. This result essentially also  holds for networks with degree correlations \cite{boguna2003absence}, although some exceptions have been pointed out \cite{eguiluz2002epidemic,blanchard2003epidemic}. In \cite{d2012robustness} the dependence of the epidemic threshold and diffusion time on the network assortativity was investigated, using a degree-preserving rewiring procedure which starts from a Barabasi-Albert network, and analysing the spectral properties of the resulting adjacency matrices. Also in this paper we mainly focus on Barabasi-Albert (BA) networks \cite{barabasi1999emergence}, using the exact results by Fotouhi and Rabbat on the degree correlations \cite{fotouhi2013degree}. We employ the mean field method \cite{vespignani2012modelling} and our network formulation of the Bass diffusion model for the description of the innovation diffusion process \cite{bertotti2016bass,bertotti2016innovation}. We solve numerically the equations and find the time of the diffusion peak, for values of the maximum network degree $n$ of the order of $10^2$, which correspond to medium-size real networks with $\simeq 10^4$ nodes. Then we compare these diffusion times with those of networks with different kinds of degree correlations.

In Section \ref{section 2}, as a preliminary to the analysis of diffusion, we compute the average nearest neighbor degree function $k_{nn}(k)$ and the Newman assortativity coefficient $r$ \cite{newman2002assortative} of BA networks. 

In Section  \ref{section 3} we write the network Bass equations and give the diffusion times found from the numerical solutions. We compare these times with those for uncorrelated scale-free networks with exponent $\gamma=3$ and with those for assortative networks whose correlation matrices are mathematically constructed and studied in another work. Then we apply a method proposed by Newman in \cite{newman2003mixing} to build disassortative correlation matrices, and evaluate the corresponding diffusion times. Our results are in qualitative agreement with those by Caldarelli et al.\ \cite{d2012robustness} for the SIS model. The minimum diffusion time is obtained for the BA networks, which have, with $n\simeq 10^2$ (corresponding to $N\simeq 10^4$ nodes, see Section \ref{discussion}), $r\simeq -0.10$. The minimum value of $r$ obtained for the disassortative networks (with $\gamma=3$) is $r\simeq -0.09$, also not far from the values obtained in Ref.\ \cite{d2012robustness}. For assortative networks, on the contrary, values of $r$ much closer to the maximum $r=1$ are easily obtained.

Section \ref{discussion} contains a discussion of the results obtained for the $r$ coefficient and the $k_{nn}$ function of BA networks.

Finally, Section \ref{Conclusions} contains our conclusions and outlook.

\section{The average nearest neighbor degree function of Barabasi-Albert (BA) networks}
\label{section 2}

We emphasize that for the networks considered in this paper a maximal value $n \in {\bf N}$
is supposed to exist for the admissible number of links emanating from each node. Notice that,
as shown later in the paper, this is definitely compatible with a high number $N$ of nodes in the network
(e.g., with a number $N$ differing from $n$ by various magnitude orders).

Together with the degree distribution $P(k)$ 
which expresses the probability that a randomly chosen node has $k$ links, 
other important statistical quantities providing information on the network structure
are the degree correlations
$P(h | k)$. Each coefficient $P(h | k)$ 
expresses the conditional probability that a node with $k$ links is connected to one with $h$ links. 
In particular, an increasing character of the average nearest neighbor degree function 
\begin{equation}
k_{nn}(k) =  \sum_{h = 1}^n h P(h | k)
\label{ANND}
\end{equation}
is a hallmark of assortative networks (i.e., of networks in which high degree nodes tend to be linked to other high degree nodes),
whereas a decreasing character of this function is to be associated to disassortative networks (networks  in which high degree nodes tend to be linked to low degree nodes).
We recall that 
the $P(k)$ and $P(h|k)$ must statisfy, besides the positivity requirements 
\begin{equation*}
P(k) \geq 0 \quad \hbox{and} \quad P(h|k) \geq 0 \, ,
\label{posit}
\end{equation*}
both the normalizations 
\begin{equation*}
\sum\limits_{k = 1}^n P(k) =1 \quad \hbox{and} \quad \sum\limits_{h = 1}^n P(h|k) =1 \, ,
\label{norm}
\end{equation*}
and the Network Closure Condition (NCC)
\begin{equation}
h P(k|h) P(h) = k P(h|k) P(k) \qquad \forall h, k  = i=1,...,n \, .
\label{NCC}
\end{equation}

A different tool usually employed to investigate 
structural properties of networks is the Newman assortativity coefficient $r$ also known as the Pearson correlation coefficient \cite{newman2002assortative}.
To define it, we need to introduce the quantities $e_{jk}$ which express the probability that a randomly chosen edge links nodes with \emph{excess degree} $j$ and $k$.
Here, the excess degree of a node is meant as its total degree minus one, namely as the number of all edges emanating from the node except the one under consideration.
The distribution of the excess degrees is easily found to be given \cite{newman2002assortative} by
\begin{equation*}
	q_k=\frac{(k+1)}{\sum_{j=1}^n j P(j)} \, P(k+1) \, .
\end{equation*}
The assortativity coefficient $r$ is then defined as 
\begin{equation}
	r = \frac{1}{\sigma^2_q} \, \sum_{k,h=0}^{n-1} kh (e_{kh}-q_kq_h) \, ,
	\label{asscoeffr}
\end{equation}
where $\sigma_q$ denotes the standard deviation of the distribution $q(k)$, i.e.,
\begin{equation*}
	\sigma^2_q = \sum_{k=1}^n k^2 q_k - \bigg(\sum_{k=1}^n k q_k \bigg)^2 \, .
\end{equation*}
The coefficient $r$ takes values in $[-1,1]$ and it owes its name to the fact that
if $r<0$, the network is disassortative, if $r=0$, the network is neutral, and if $r>0$, the network is assortative.
A formula expressing the $e_{kh}$ in terms of known quantities is necessary if one wants to calculate $r$. This can be obtained 
as discussed next.
Let us move from the elements $E_{kh}$ expressing the number of edges which link
nodes with degree $k$ and $h$, with the only exception that 
edges linking nodes with the same degree have to be counted twice \cite{newman2003structure,barabasi2016network}. 
Define now $\tilde e_{kh} = E_{kh}/ (\sum_{k,h} E_{kh})$.
Each $\tilde e_{kh}$ corresponds then to the fraction of edges linking
nodes with degree $k$ and $h$ (with the mentioned interpretation of the $\tilde e_{kk}$).
We also observe that $e_{k,h} = \tilde e_{k+1,h+1}$ holds true.
The degree correlations can be related to the $\tilde e_{kh}$ through the formula 
\begin{equation}
P(h | k) = \frac{\tilde e_{kh}}{\sum_{j=1}^n \tilde e_{kj}} \qquad \forall h, k  = i=1,...,n \, .
\label{Phkintermsofehk}
\end{equation}
What is of interest for us here is the following ``inverse'' formula:
\begin{equation}
\tilde e_{hk} = \frac{P(h|k) k P(k)}{\sum_{j=1}^n j P(j)} \, .
\label{ehkintermsofPhk}
\end{equation}

\bigskip

In the rest of this section we discuss the quantities which allow to calculate the average nearest neighbor degree function $k_{nn}(k)$ and the coefficient $r$ 
for finite Barabasi-Albert (BA) networks.

The degree distribution of the Barabasi-Albert networks is known to be given by
\begin{equation}
	P(k) = \frac{2 \beta (\beta + 1)}{k (k+1) (k+2)} \, ,
	\label{PkBA}
\end{equation}
where $\beta \ge 1$ is the 
parameter in the preferential attachment procedure characterizing them \cite{barabasi1999emergence,barabasi2016network}. 
In particular, $(\ref{PkBA})$ yields $P(k) \sim c/k^3$ with a suitable constant $c$ for large $k$.

An explicit expression for the degree correlations $P(h | k)$ was given by Fotouhi and Rabbat in \cite{fotouhi2013degree}. They showed that for a growing network in the asymptotic limit
as $t \to \infty$,
\begin{equation}
	P(h | k) = \frac{\beta}{kh} \bigg( \frac{k+2}{h+1} - B^{2\beta+2}_{\beta+1} \, \frac{B^{k+h-2\beta}_{h-\beta}}{B^{k+h+2}_h} \bigg) \, ,
	\label{PhkBA}
\end{equation}
with $B^m_j$ denoting the binomial coefficient 
\begin{equation*}
	B^m_j=\frac{m!}{j! (m-j)!} \, .
\end{equation*}
Since the networks we consider have a maximal number of links $n$ \cite{pastor2002epidemic}, we must normalize the matrix with elements $P(h | k)$.
We calculate $C_k = \sum_{h = 1}^n P(h|k)$ and take as a new degree correlation matrix the matrix whose $(h,k)$-element is 
\begin{equation}
	P_n(h | k) = \frac{P(h | k)}{C_k} \, ,
	\label{PhkfiniteBA}
\end{equation}
with $P(h | k)$ as in $(\ref{PhkBA})$.

The average nearest neighbor degree function $k_{nn}(k)$ and the coefficient $r$ can be now easily calculated with a software like \texttt{Mathematica},
by using $(\ref{ANND})$, $(\ref{asscoeffr})$, $(\ref{PkBA})$ and $(\ref{PhkfiniteBA})$.
Results are reported and discussed in Section \ref{discussion}.

\section{The Bass diffusion equation on complex networks}
\label{section 3}

In \cite{bertotti2016bass,bertotti2016innovation} we have reformulated the well-known Bass equation of innovation diffusion
\begin{equation}
\frac{{d{F}(t)}}{{dt}} = [1 - F(t)]\left[ {p + qF(t)} \right] 
\label{Bass}
\end{equation}
(where $F(t)$ is the cumulative adopter fraction at the time $t$, and $p$ and $q$ are the innovation and the imitation coefficient respectively),
providing versions suitable
for the case in which the innovation diffusion process occurs on a network.
The model can be expressed in such a case by
a system of $n$ ordinary differential equations, $n$ being the maximal number of links of a node of the network, 
\begin{equation}
\frac{{d{G_i}(t)}}{{dt}} = [1 - {G_i}(t)]\left[ {p + iq\sum\limits_{h = 1}^n {P(h|i)\,{G_h}(t)} } \right] \qquad i=1,...,n \, .
\label{BassNethomo}
\end{equation}
The quantity $G_i(t)$ in (\ref{BassNethomo}) represents for any $i = 1, ... n$ the fraction of potential adopters with $i$ links that at the time $t$ have adopted the innovation.
More precisely, denoting by $F_i(t)$ the fraction of the total population composed by individuals with $i$ links, who at the time $t$ have adopted,
we set $G_i(t)=F_i(t)/P(i)$.
Further heterogeneity can be introduced allowing also the innovation 
coefficient (sometimes called publicity coefficient) to be dependent on $i$. In this case, the equations take the form
\begin{equation}
\frac{{d{G_i}(t)}}{{dt}} = [1 - {G_i}(t)]\left[ {p_i + iq\sum\limits_{h = 1}^n {P(h|i)\,{G_h}(t)} } \right] \qquad i=1,...,n \, , 
\label{BassNethetero}
\end{equation}
and, for example, 
the $p_i$ can be chosen to be inversely proportional to $P(i)$ or to have a linear dependence, decreasing in $i$; in the first case more publicity is delivered to the hubs, with ensuing ``trickle-down'' diffusion, while in the second case a ``trickle-up'' diffusion from the periphery of the network can be simulated. 

\begin{figure}[h]
  \begin{center}
\includegraphics[width=7.0cm,height=4.1cm]{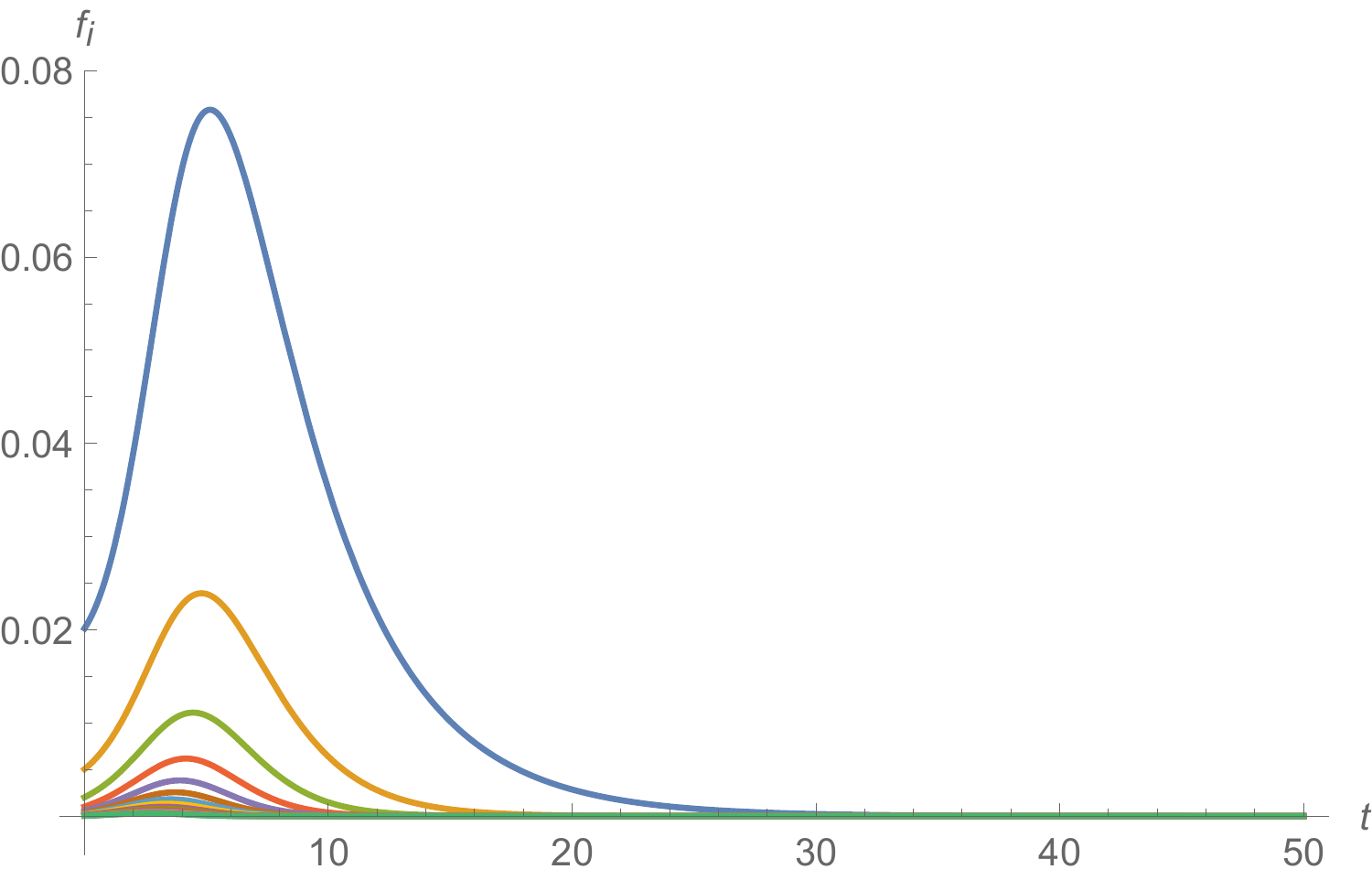}
  \hskip0.7cm
\includegraphics[width=7.0cm,height=4.1cm]{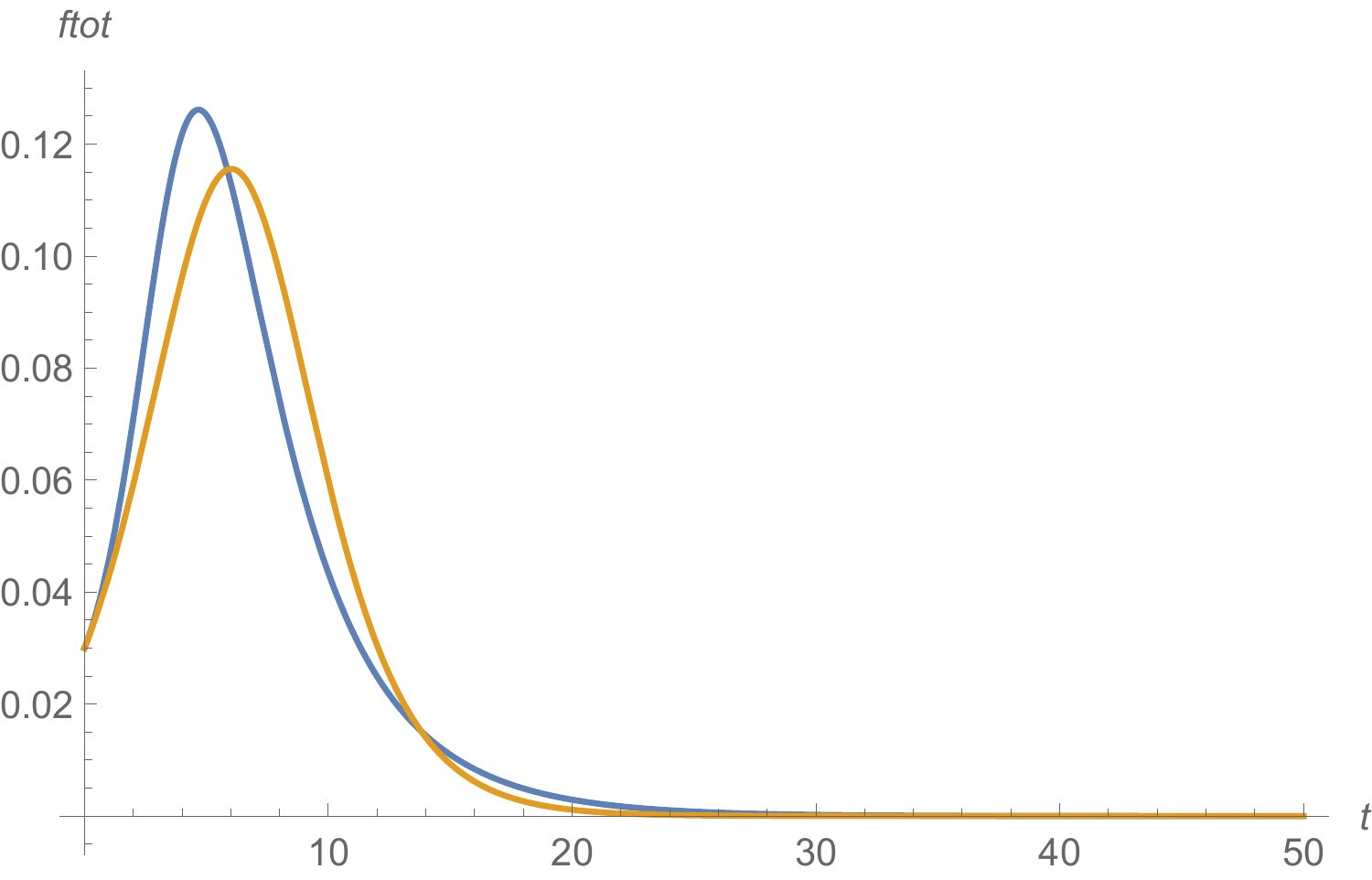}
\caption{\textbf{Left panel:} Fraction $f_i$ of new adoptions per unit time in the ``link class $i$'' (the subset of all individuals having $i$ links), as a function of time, in the Bass innovation diffusion model on a BA network. The parameter $\beta$ of the network (number of child nodes in the preferential attachment scheme) is $\beta=1$. The maximum number of links in this example is $n=15$, while in the comprehensive numerical solutions, whose results are summarized in Fig.\ \ref{fig1}, it is $n=99$. The adoption peak occurs later for the least connected (and most populated) class with $i=1$.
\textbf{Right panel:} Cumulative new adoptions $f_{tot}=\sum_{i=1}^n f_i$ per unit time, as a function of time, compared with the same quantity for the homogeneous Bass model without an underlying network. The peak of the homogeneous Bass model is slightly lower and shifted to the right. For the values of the model parameters $p$, $q$ and the measuring unit of time see Fig.\ \ref{fig1}.
} 
\label{figA}
  \end{center}
\end{figure}

The function $f_i(t)=\dot{F}_i(t)$ gives the fraction of new adoptions per unit time in the ``link class $i$'' (or ``degree class $i$'') i.e., in the subset of individuals having $i$ links. 
The left panel in Fig.\ \ref{figA} 
shows an example of a numerical solution with plots of all the $f_i$'s, in a case where for graphical reasons we have taken $n$ small ($n=15$). The underlying network is a BA with $\beta=1$. As is clear from the plot, the largest fraction of new adopters belongs at all times to the link class with $i=1$, which reaches its adoption peak later than the others. In general, the more connected individuals are, the earlier they adopt. This phenomenon is quite intuitive and has been evidenced in previous works on mean-field epidemic models, see for instance \cite{barthelemy2005dynamical}. As discussed in our paper \cite{bertotti2016innovation}, in applications of the Bass model to marketing this may allow to estimate the $q$ coefficient, when it is not known in advance, by monitoring the early occurrence of adoption in the most connected classes. 
The right panel in Fig.\ \ref{figA} 
shows the total adoption rate $f(t)$ corresponding to the same case as in 
the left panel.
The simple Bass curve (homogeneous model, without underlying network) is also shown for comparison.

In the Bass model, unlike in other epidemic models where infected individuals can return to the susceptible state, the diffusion process always reaches all the population. The function $f(t)=\sum\limits_{i = 1}^n\dot{F}_i(t)$, which represents the total number of new adoptions per unit time, usually has a peak, as we have seen in the previous example. We choose the time of this peak as a measure of the diffusion time; it is computed for each numerical solution of the diffusion equations by sampling the function $f(t)$.  For fixed coefficients of publicity $p$ and imitation $q$, the time depends on the features of the network. In this paper we consider only scale-free networks with $\gamma=3$, for comparison with BA networks. 

\begin{figure}
\begin{center}
\includegraphics[width=12cm,height=7.8cm]{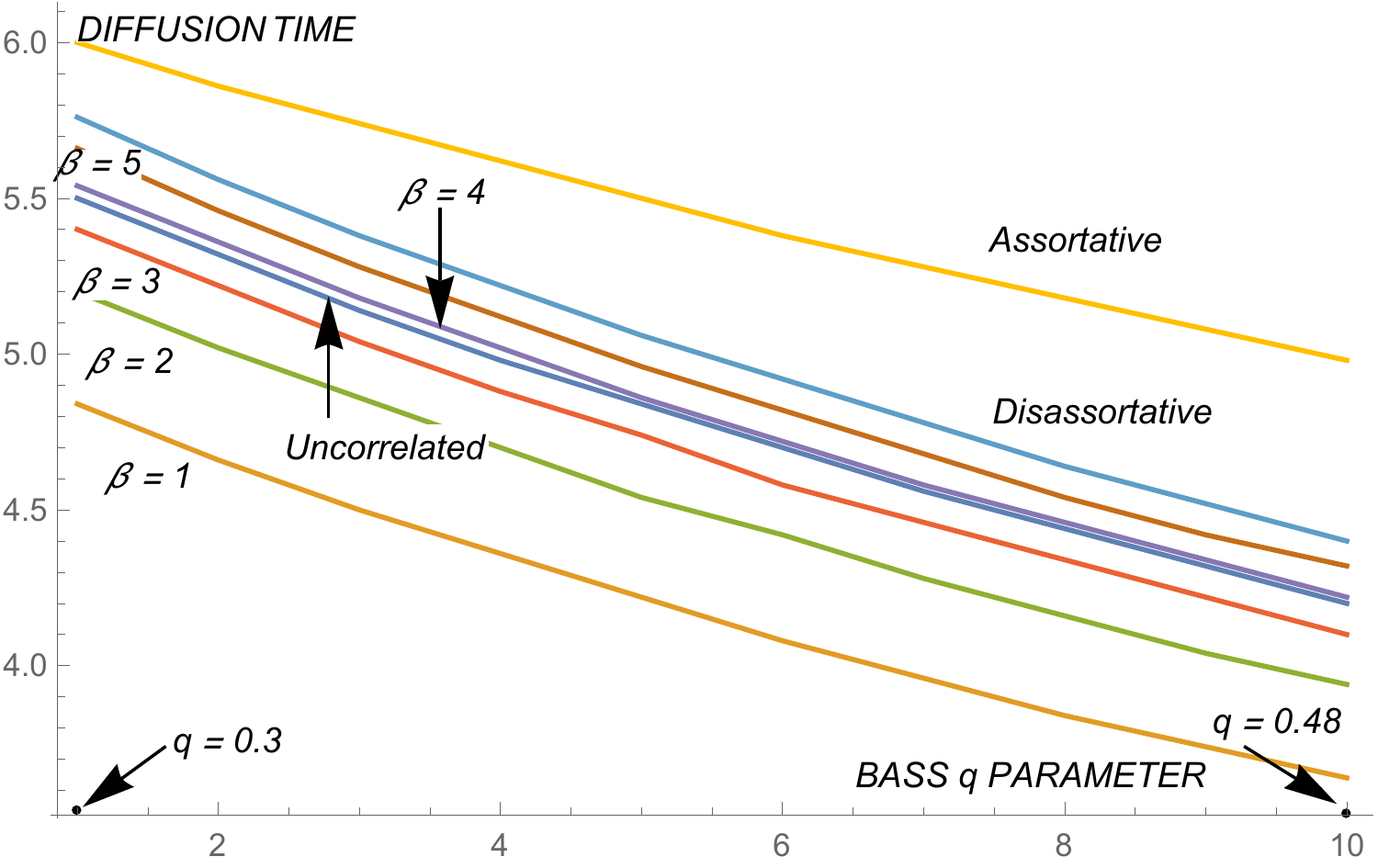}
\caption{Time of the diffusion peak (in years) for the Bass model on different kinds of scale-free networks with exponent $\gamma=3$, as a function of the imitation coefficient $q$. All networks have maximum degree $n=99$. The $q$ coefficient varies in the range $0.3 - 0.48$, corresponding to a typical set of realistic values in innovation diffusion theory \cite{jiang2006virtual}. 
The publicity coefficient is set to the value $p=0.03$ (see Fig.\ \ref{figPVAR} for results with an inhomogeneous $p_k$ depending on the link class $k=1, \ldots ,n$).
The lines with $\beta=1,2,3,4,5$ correspond to BA networks with those values of $\beta$. Their assortativity coefficients are respectively $r=-0.104,-0.089, -0.078,-0.071,-0.065$. The disassortative network is built with the Newman recipe (Sect.\ \ref{matriciD}) with $d=4$ and has $r=-0.084$. The assortative network is built with our recipe (Sect.\ \ref{matriciA}), with $\alpha=1/2$, and has $r=0.863$.
} 
\label{fig1}
\end{center}  
\end{figure}

Fig.\ \ref{fig1} shows the peak times obtained for different networks with maximum degree $n=99$, as a function of the imitation parameter $q$. This value of $n$ has been chosen because it corresponds to a number $N$ of nodes of the order of $10^4$; this allows a comparison with the results of Caldarelli et al.\ (see below) and displays finite-size effects, as discussed in Sect.\ \ref{discussion} (note, however, that such effects are still present with $N \simeq 10^6$ and larger). The choice of an odd value for $n$ is necessary for the construction procedure of the assortative matrices (Sect.\ \ref{matriciA}).

It is clear from Fig.\ \ref{fig1} that diffusion is faster on the BA networks with $\beta=1,2,3$ than on an uncorrelated network. For $\beta=4$ the diffusion time is almost the same, and for $\beta=5$ (and $\beta >5$, not shown) diffusion on the BA network is slower than on the uncorrelated network. On purely disassortative networks diffusion is slightly slower than on an uncorrelated network, and much slower on assortative networks.

\subsection{Comparison with the SIS model}

In \cite{d2012robustness} Caldarelli et al.\ have studied the dependence of the epidemic threshold and diffusion time for the SIS epidemic model on the assortative or disassortative character of the underlying network. Both the epidemic threshold and the diffusion time were evaluated from the eigenvalues of the adjacency matrix. The networks employed had a number of nodes $N=10^4$, a scale-free degree distribution with exponent $\gamma=3$, and were obtained from a BA seed network through a Monte Carlo rewiring procedure which preserves the degree distribution but changes the degree correlations. The Monte Carlo algorithm employs an ``Hamiltonian'' function which is related to the Newman correlation coefficient $r$. The values of $r$ explored in this way lie in the range from -0.15 to 0.5 and are therefore comparable with those obtained for our assortative and disassortative matrices.

Although the epidemic model considered and the definition of the diffusion time adopted in \cite{d2012robustness} are different from ours, there is a qualitative agreement in the conclusions: the diffusion time increases with the assortativity of the network, and is at a minimum for values of $r$ approximately equal to -0.10. This value of $r$ corresponds to those of finite BA networks with $\beta=1$ and $N \simeq 10^4$ and is slightly smaller than the minimum value of $r$ obtained for disassortative networks with a matrix $e_{jk}$ built according to Newman's recipe (Sect.\ \ref{matriciD}). Note that for those networks the function $k_{nn}(k)$ is decreasing for any $k$, while for BA networks it is decreasing at small values of $k$ and increasing at large $k$ (Sect.\ \ref{discussion}); nevertheless, their $r$ coefficient never becomes significantly less than $\simeq -0.1$, even with other choices in the recipe, as long as $\gamma=3$. 
We also recall that the clustering coefficient of the BA networks is exactly zero for $\beta=1$; for an analysis of the role of the clustering coefficient in epidemic spreading on networks see for instance \cite{isham2011spread}.

A re-wiring procedure comparable to that of Ref.\ \cite{d2012robustness} (without evaluation of the epidemic threshold and diffusion time) has been mathematically described by Van Mieghem et al.\ \cite{van2010influence}.

In the next subsections we 
provide information on the degree correlation matrices and other features of
the networks used above for comparison with the BA networks.
To compare diffusion peak times on networks which, although characterized by
different assortativity/disassortativity properties, share some similarity, we consider
networks whose degree distribution $P(k)$ obeys a power-law with exponent three, i.e. is of the form 
\begin{equation}
P(k) = \frac{c}{k^3} \, ,
\label{powerlaw3}
\end{equation}
where $c$ is the normalization constant. 

\subsection{Uncorrelated networks}
\label{uncorrelated}

Let us start with uncorrelated networks. As their name suggests,
in these networks the degree correlations $P(h|k)$ do not depend on $k$. They can be easily seen to be given by 
\begin{equation}
	P(h | k) = \frac{h P(h)}{\langle h \rangle} \, ,
\label{Phkuncorr}
\end{equation}
with $\langle h \rangle = \sum_{h = 1}^n {h P(h)}$, and hence
their average nearest neighbor degree function $(\ref{ANND})$ is 
\begin{equation}
k_{nn}(k) =  \frac{\langle h^2 \rangle}{\langle h \rangle} \, ,
\end{equation}
a constant. The coefficient $r$ is trivially found to be equal to zero.

\subsection{A family of assortative networks}
\label{matriciA}

\begin{figure}
\begin{center}
\includegraphics[width=7cm,height=4.5cm]{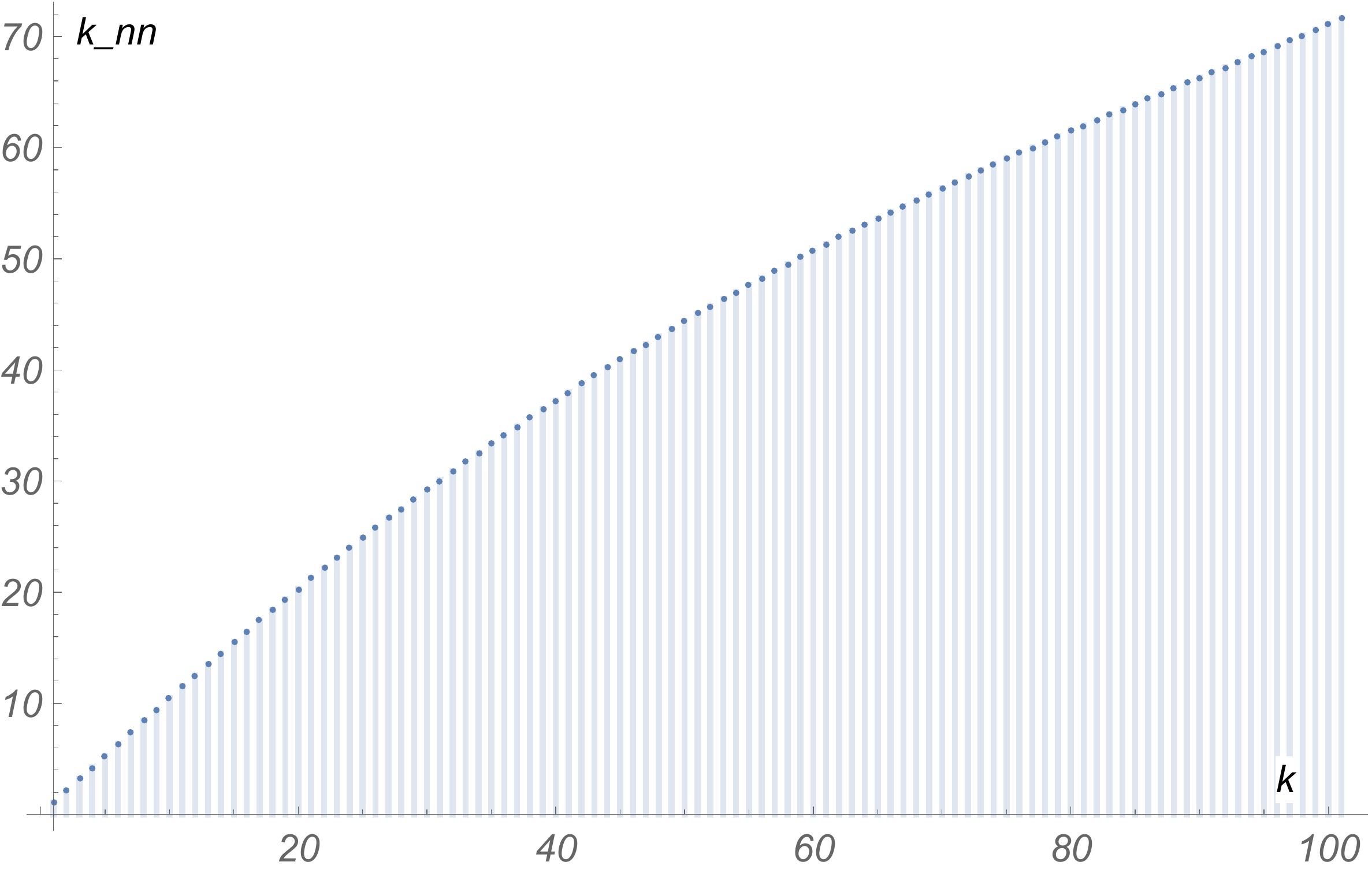}
\caption{Function $k_{nn}$ for an assortative network as in Sect.\ \ref{matriciA} with $\alpha=1/2$, $n=101$. 
} 
\label{figknnkass}
\end{center}  
\end{figure}

We consider now a family of assortative networks we have introduced in \cite{bertotti2016bass}.
To give here the expressions of their degree correlations $P(h|k)$, we need to recall their construction.
We start defining the elements of a $n \times n$ matrix $P_0$ as 
\begin{equation}
{P_0}(h|k) = |h - k|^{- \alpha} \quad \hbox{if} \ h < k \quad \hbox{and} \quad {P_0}(h|k) =1 \quad \hbox{if} \ h = k \, ,
\label{e15}
\end{equation}
for some parameter $\alpha > 0$,
and we define the elements ${P_0}(h|k)$ with $h>k$ in such a way that the formula (\ref{NCC}) is satisfied by the ${P_0}(h|k)$.
Hence, since the normalization $\sum_{h = 1}^n P_0(h|k) =1$ has to hold true, we compute for any $k = 1,...,n$ the sum
$C_k = \sum_{h = 1}^n P_0(h|k)$ and call $C_{max} = \max_{k=1,...,n} \, C_k $.
Then, we introduce a new matrix $P_1$, requiring that its diagonal elements be given by
${P}_1(k|k)= C_{max} - C_k$ for any $k = 1,...,n$,
whereas the non-diagonal elements are the same as those of the matrix $P_0$: ${P}_1(h|k)={P}_0(h|k)$ for $h \ne k$.
For any $k=1,...,n$ the column sum $\sum_{h = 1}^n {P}_1(h|k)$ is then equal to $C_k - 1 + C_{max} - C_k = C_{max} - 1.$
Finally, we normalize the entire matrix by setting
\begin{equation}
{P}(h|k) = \frac{1}{(C_{max} - 1)} \,  {P}_1(h|k) \qquad \hbox{for} \ h,k = 1,...,n \, .
\label{Phkassort}
\end{equation}
Again, the average nearest neighbor degree function $k_{nn}(k)$ and the coefficient $r$ can be calculated with a software.
The increasing character of $k_{nn}(k)$ for a network of the family in this subsection with $\alpha = 1/2$ and $n=101$
is shown for example in Fig.\ \ref{figknnkass}.

\subsection{A family of disassortative networks}
\label{matriciD}

\begin{figure}
\begin{center}
\includegraphics[width=7cm,height=4.5cm]{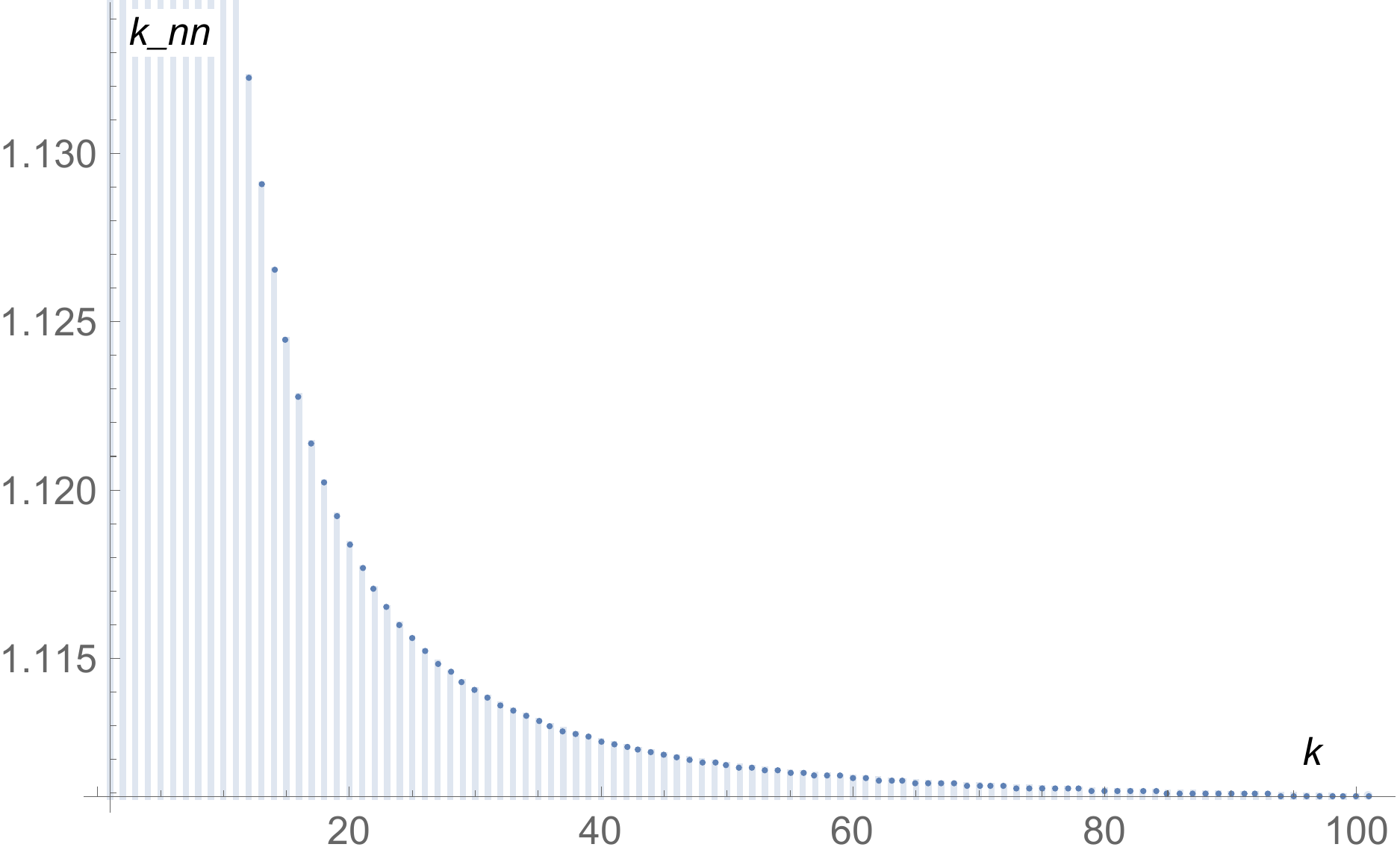}
\caption{Function $k_{nn}$ for a disassortative network as in Sect.\ \ref{matriciD} with $\gamma=4$, $n=101$. 
} 
\label{figknnkdisass}
\end{center}  
\end{figure}

Different models of disassortative networks can be constructed based on a suggestion by Newman contained in \cite{newman2003mixing}.
According to it, one can set $e_{kh} = q_k x_h + x_k q_h - x_k x_h$
for $h, k  = 0,...,n-1$,
where $q_k$ is the distribution of the excess degrees and $x_k$ is any distribution satisfying $\sum_{k=0}^{n-1} x_k = 1$, and with 
$x_k$ decaying faster than $q_k$.
Choosing, to fix ideas, $x_k = (k+1)^{-\gamma}/\sum_{j=0}^{n-1} (j+1)^{-\gamma}$ with a parameter $\gamma > 2$, 
we denote $S = \sum_{j=0}^{n-1} (j+1)^{-\gamma}$ and $T =  \sum_{j=0}^{n-1} (j+1)^{-2}$ 
and then set for all $h, k  = 0,...,n-1$,
\begin{equation}
e_{kh} = \bigg( \frac{1}{ST}\Big((k+1)^{-2} (h+1)^{-\gamma} + (h+1)^{-2} (k+1)^{-\gamma}\Big) - \frac{1}{S^2}(k+1)^{-\gamma} (h+1)^{-\gamma} \bigg) \, .
\label{ehkNewmandis}
\end{equation}
We show in the Appendix that the inequalities $0 \le e_{kh} \le 1$ hold true for any $h, k  = 0,...,n-1$.
In view of $(\ref{Phkintermsofehk})$, the degree correlations are then obtained as  
\begin{equation}
P(h | k) = \frac{e_{k-1,h-1}}{\sum_{j=1}^n e_{k-1,j-1}} \, , \qquad \forall h, k  = 1,...,n \, ,
\end{equation}
with the $e_{kh}$ as in $(\ref{ehkNewmandis})$.
It is immediate to check that the coefficient $r$ is negative, see e.g. \cite{newman2003mixing}.
As for the average nearest neighbor degree function $k_{nn}(k)$, it can be calculated with a software. 
The decreasing character of $k_{nn}(k)$ for a network of the family in this subsection with $\gamma = 4$ and $n=101$
is shown for example in Fig.\ \ref{figknnkdisass}.

\bigskip

\begin{figure}
\begin{center}
\includegraphics[width=8cm,height=5.2cm]{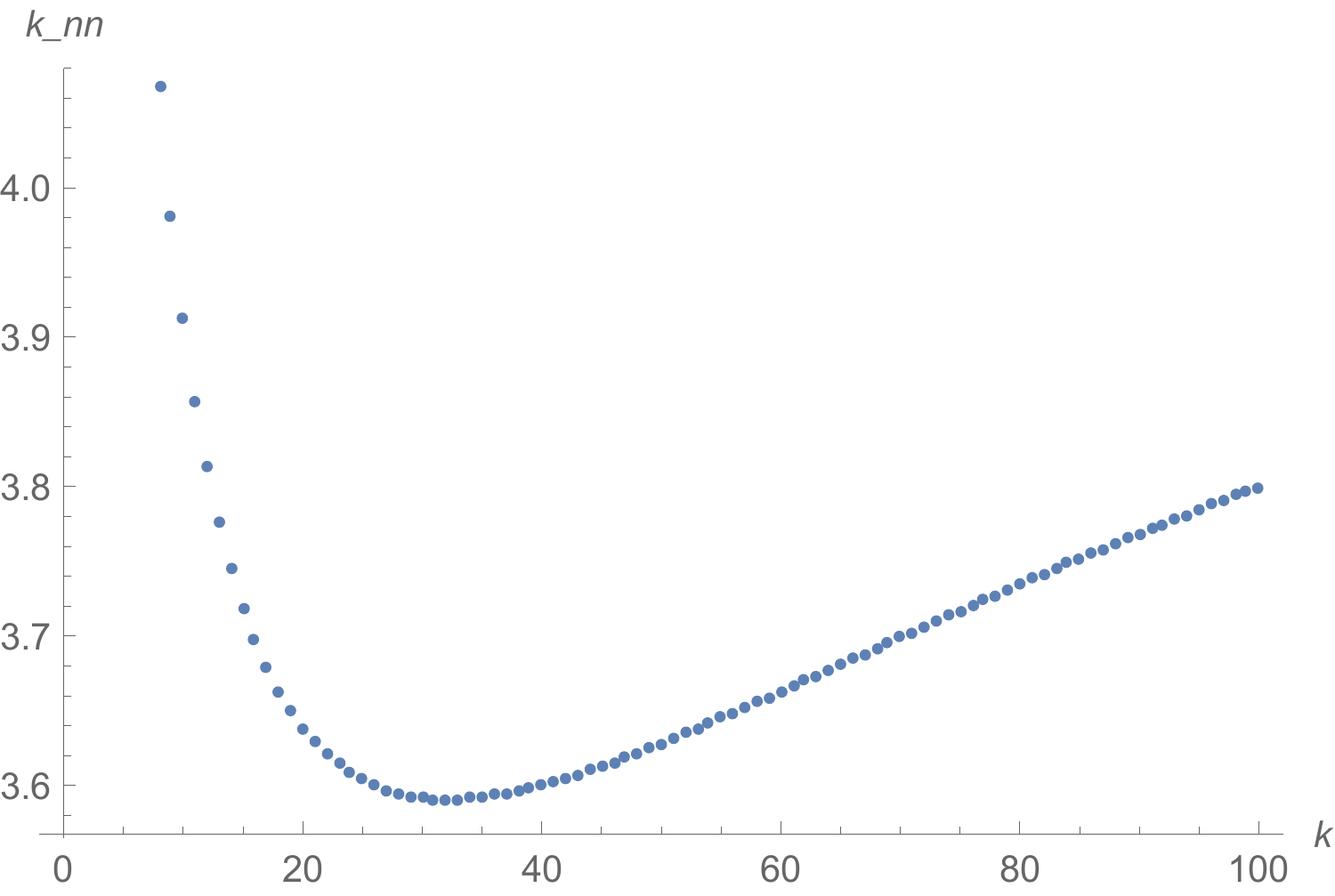}
\caption{Function $k_{nn}$ for a BA network with $\beta=1$, $n=100$ (largest degree). 
} 
\label{figK1}
\end{center}  
\end{figure}

\begin{figure}
\begin{center}
\includegraphics[width=8cm,height=5.2cm]{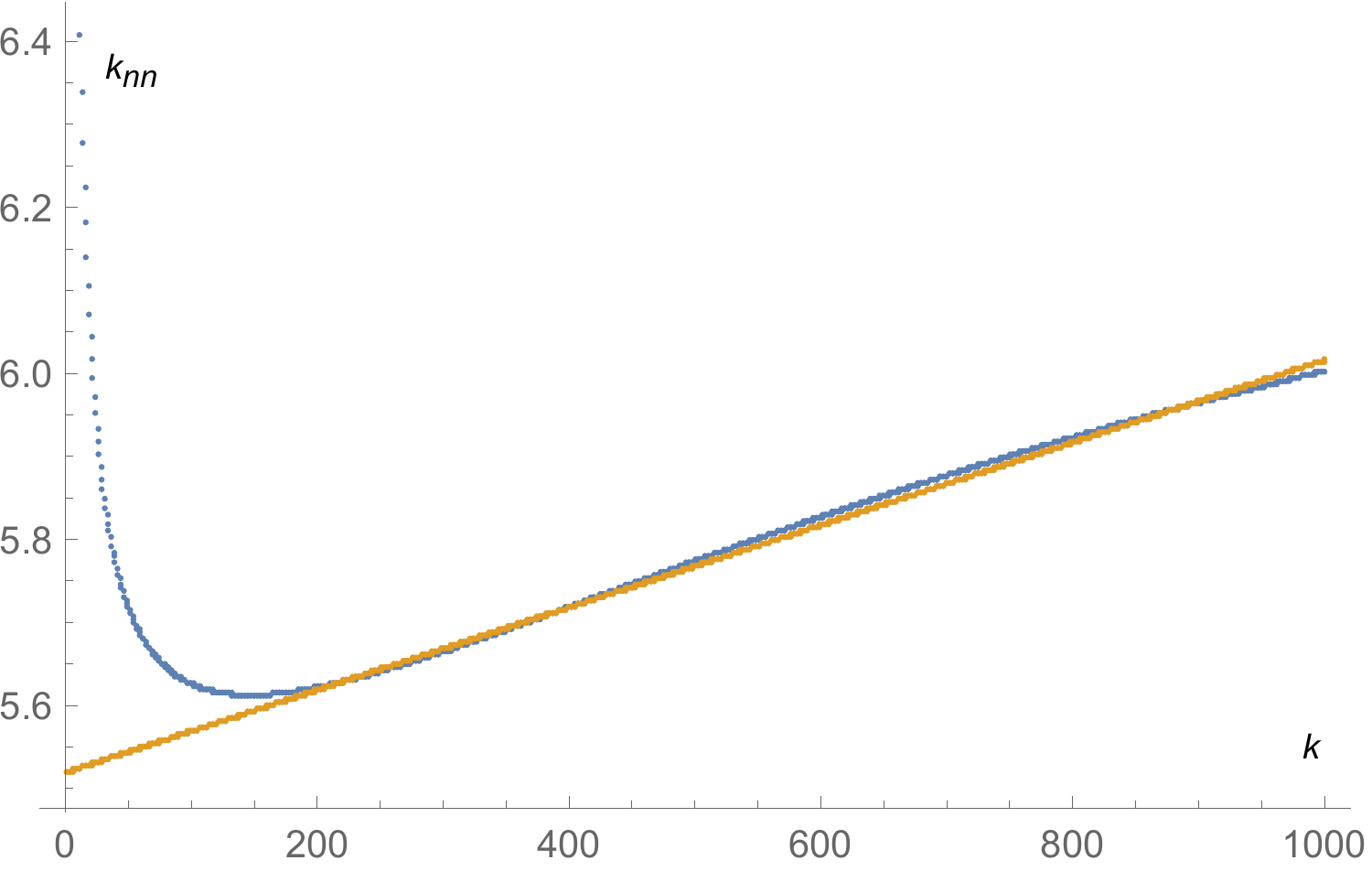}
\caption{Function $k_{nn}$ for a BA network with $\beta=1$, $n=1000$ (largest degree). A linear fit for large $k$ is also shown. The slope is approximately equal to $5\cdot 10^{-4}$; a comparison with Fig.\ \ref{figK1} shows that the slope decreases with increasing $n$, but is still not negligible.
} 
\label{figK2}
\end{center}  
\end{figure}

\begin{figure}
\begin{center}
\includegraphics[width=8cm,height=5.2cm]{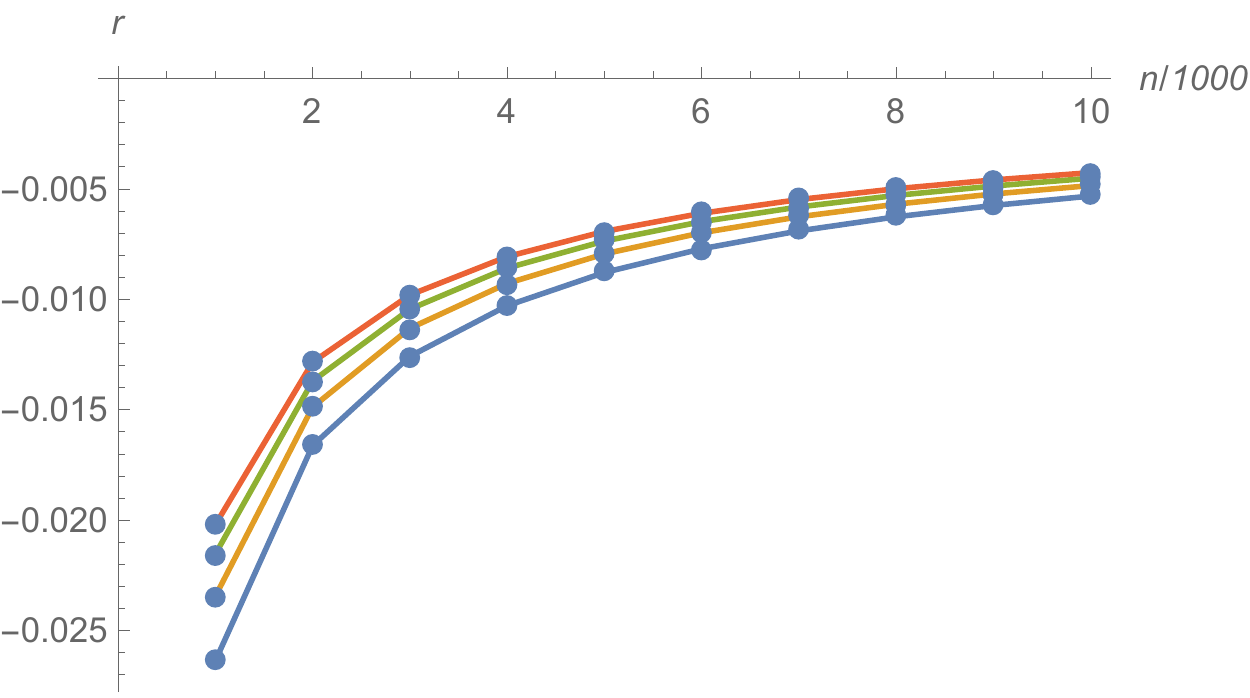}
\caption{Newman assortativity coefficient $r$ for BA networks with $\beta=1,2,3,4$ and largest degree $n=1000, \ 2000, ... , 10000$.
} 
\label{fig5}
\end{center}  
\end{figure}

\section{Discussion}
\label{discussion}

\subsection{The Newman assortativity coefficient for BA networks}

In \cite{newman2002assortative} Newman reported in a brief table the values of $r$ for some real networks. More data are given in his book \cite{newman2010networks}. Focusing on disassortative scale-free networks for which the scale-free exponent $\gamma$ is available, one realizes that their negative $r$ coefficient is generally small in absolute value, especially when $\gamma$ is equal or close to 3. For instance:

\begin{itemize}

\item WWW \texttt{nd.edu}: $\gamma$ from $2.1$ to $2.4$, $r=-0.067$

\item Internet: $\gamma=2.5$, $r=-0.189$

\item Electronic circuits: $\gamma=3.0$, $r=-0.154$

\end{itemize}

The size of these networks varies, being of the magnitude order of $N \approx 10^4$ to $N \approx 10^6$. The data mentioned above have been probably updated and extended, but in general it seems that scale-free disassortative networks with these values of $\gamma$ tend to have an $r$ coefficient that is much closer to 0 than to $-1$. As we have seen, this also happens with disassortative scale-free networks mathematically defined, whose degree correlations are given by a procedure also introduced by Newman.

For ideal BA networks, Newman gave in  \cite{newman2002assortative} an asymptotic estimate of the $r$ coefficient based on the correlations computed in \cite{krapivsky2001organization} and concluded that $r$ is negative but vanishes as $(\log N)^2/N$ in the large-$N$ limit. Other authors share the view that BA networks are essentially uncorrelated \cite{noldus2015assortativity,barthelemy2005dynamical}. The smallness of their $r$ coefficient is also confirmed by numerical simulations, in which the network is grown according to the preferential attachment scheme, even though the results of such simulations are affected by sizable statistical errors when $N$ varies between $\approx 10^4$ and $\approx 10^6$.

In the recent paper \cite{fotouhi2018temporal} Fotouhi and Rabbat use their own expressions for the $p(l,k)$ correlations 
(expressing the fraction of links whose incident nodes have degrees $l$ and $k$)
to compute the asymptotic behavior of $r$ according to an alternative expression given 
by Dorogovtsev and Mendes \cite{dorogovtsev2002evolution}. They conclude that the estimate $r \simeq (\log N)^2/N$ given by Newman is not correct. They find $|r|\approx (\log n)^2/n$. In order to relate the network size $N$ to the largest degree $n$, they use the relation $n \simeq \sqrt{N}$ based on the continuum-like criterium $\int_n^\infty P(k)dk=N^{-1}$ \cite{boguna2004cut}. So in the end $|r|\approx (\log N)^2/\sqrt{N}$. They check that this is indeed the correct behavior by performing new large-scale simulations.

We have computed $r$ using the original definition by Newman \cite{newman2002assortative}, based on the matrix $e_{jk}$, and the exact $P(h|k)$ coefficients, with $n$ up to 15000, which corresponds to $N \simeq 225000000$ (Fig.\ \ref{fig5}). We found a good agreement with the estimate of Fotouhi and Rabbat. Note, however, that although the ``thermodinamic'' limit of $r$ for infinite $N$ is zero, for finite $N$ the value of $r$ cannot be regarded as vanishingly small, in consideration of what we have seen above for disassortative scale-free networks with $\gamma=3$.

\subsection{The function $k_{nn}(k)$}

An early estimate of the function $k_{nn}(k)$ valid also for BA networks has been given by Vespignani and Pastor-Satorras in an Appendix of their book on the structure of the Internet \cite{pastor2007evolution}. Their formula is based in turn on estimates of the behavior of the conditional probability
$P(k'|k)$ which hold for a growing scale-free network with a degree distribution $P(k) \propto k^{-\gamma}$. This formula reads
\begin{equation} 
P(k'|k) \propto k'^{-(\gamma-1) } k^{-(3-\gamma) }
\label{pastor}
\end{equation} 
and holds under the condition $1 \ll k' \ll k$.
Using the exact $P(k'|k)$ coefficients of Fotouhi and Rabbat it is possible to check the approximate validity of this formula. 

In Ref.\ \cite{pastor2007evolution} the expression (\ref{pastor}) is then used to estimate $k_{nn}(k)$. Since, for $\gamma=3$, $P(k'|k)$ does not depend on $k$, the conclusion is that $k_{nn}$ is also independent from $k$. It is not entirely clear, however, how the condition $1 \ll k' \ll k$ can be respected when the sum over $k'$ is performed, and how the diverging factor $\sum_{k'=1}^n (1/k') \simeq \ln (n) \simeq \frac{1}{2} \ln (N)$ should be treated.

In their recent work \cite{fotouhi2018temporal}, Fotouhi and Rabbat use their own results  for the $P(k'|k)$ coefficients in order to estimate $k_{nn}(k)$ in the limit of large $k$. They find $k_{nn}(k) \approx \beta \ln (n)$ and cite a previous work \cite{pastor2005rate} which gives the same result, in the form $k_{nn}(k) \approx \frac{1}{2} \beta \ln (N)$ (we recall that $n \simeq \sqrt{N}$). In this estimate we can observe, as compared to Ref.\ \cite{pastor2007evolution}, the explicit presence of the parameter $\beta$ and the diverging factor $\ln(n)$.

Concerning the Newman assortativity coefficient $r$, Fotouhi and Rabbat also make clear that even though $r \to 0$ when $N \to \infty$, this does not imply that the BA networks are uncorrelated, and in fact the relation 
$k_{nn}(k)=\langle k^2 \rangle / \langle k \rangle$, valid for uncorrelated networks, does not apply to BA networks. 

A direct numerical evaluation for finite $n$ of the function $k_{nn}(k)$ based on the exact $P(k'|k)$ shows further interesting features. As can be seen in Figs.\ \ref{figK1}, \ref{figK2}, the function is decreasing at small $k$ and slightly and almost linearly increasing at large $k$. Note that this happens for networks of medium size ($n=100$, $N \simeq 10^4$) like those employed in our numerical solution of the Bass diffusion equations and employed in Ref.\ \cite{d2012robustness}, but also for larger networks (for instance with $n=1000$, $N \simeq 10^6$, compare Fig.\ \ref{figK2}). It seems that the ``periphery'' of the network, made of the least connected nodes, has a markedly disassortative character, while the hubs are slightly assortative. (On general grounds one would instead predict for finite scale-free networks a small structural disassortativity at large $k$ \cite{barabasi2016network}.) 

Since evidence on epidemic diffusion obtained so far indicates that generally the assortative character of a network lowers the epidemic threshold and the disassortative character tends to make diffusion faster once it has started, this ``mixed'' character of the finite BA networks appears to facilitate spreading phenomena and is consistent with our data on diffusion time (Sect.\ \ref{section 3}). Note in this connection that some real networks also turn out to be both assortative and disassortative, in different ranges of the degree $k$; compare the examples in \cite{barabasi2016network}, Ch.\ 7.

Finally, we would like to relate our numerical findings for the function $k_{nn}$ to the general property (compare for instance \cite{boguna2003absence})
\begin{equation}
	\langle k^2 \rangle = \sum_{k=1}^n k P(k) K(k,n),
	\label{eqA}
\end{equation}
where for simplicity $K$ denotes the function $k_{nn}$ and the dependence of this function on the maximum degree $n$ is explicitly shown. For BA networks with $\beta=1$ we have at large $n$ on the l.h.s.\ of (\ref{eqA}), from the definition of $\langle k^2 \rangle$,
\begin{equation}
	\langle k^2 \rangle = \sum_{k=1}^n k^2 \frac{4}{k(k+1)(k+2)} = 4\ln(n) + o(n),
	\label{eqB}
\end{equation}
where the symbol $o(n)$ denotes terms which are constant or do not diverge in $n$.

For the expression on the r.h.s.\ of (\ref{eqA}) we obtain 
\begin{equation}
	\sum_{k=1}^n k P(k) K(k,n) = \sum_{k=1}^n \frac{4}{(k+1)(k+2)} K(k,n)
	\label{eqC}
\end{equation}

Eqs.\ (\ref{eqB}) and (\ref{eqC}) are compatible, in the sense that their diverging part in $n$ is the same, in two cases: (1) if for large $k$ we have $K(k,n) \simeq a \ln(n)$, where $a$ is a constant; 
this is true because in that case the sum on $k$ is convergent; (2) if more generally $K(k,n) \simeq a \ln(n) + b(n)k$; this is still true because also for the term $b(n)k$ the sum in $k$ 
leads to a result proportional to $\ln(n)$. Case (2) appears to be what happens, according to Figs.\ \ref{figK1}, \ref{figK2}.

\section{Diffusion times with heterogeneous $p$ coefficients}
\label{heterog}

In the Bass model the publicity coefficient $p$ gives the adoption probability per unit time of an individual who has not yet adopted the innovation, independently from the fraction of current adopters. Therefore, it is not due to the word-of-mouth effect, but rather to an external stimulus (advertising) which is received in the same way by all individuals. The intensity of this stimulus is in turn proportional to the advertising expenses of the producer or seller of the innovation. One can therefore imagine the following alternative to the uniform dissemination of ads to all the population: the producer or seller invests for advertising to each individual in inverse proportion to the population of the individual's link class, so as to speed up adoption in the most connected classes and keep the total expenses unchanged. In terms of the $p_i$ coefficients in eq.\ (\ref{BassNethetero}) this implies $p_i \propto 1/P(i)$, with normalization $\sum_{i=1}^n p_iP(i)=p$ \cite{bertotti2016bass}.

\begin{figure}
\begin{center}
\includegraphics[width=12cm,height=7.8cm]{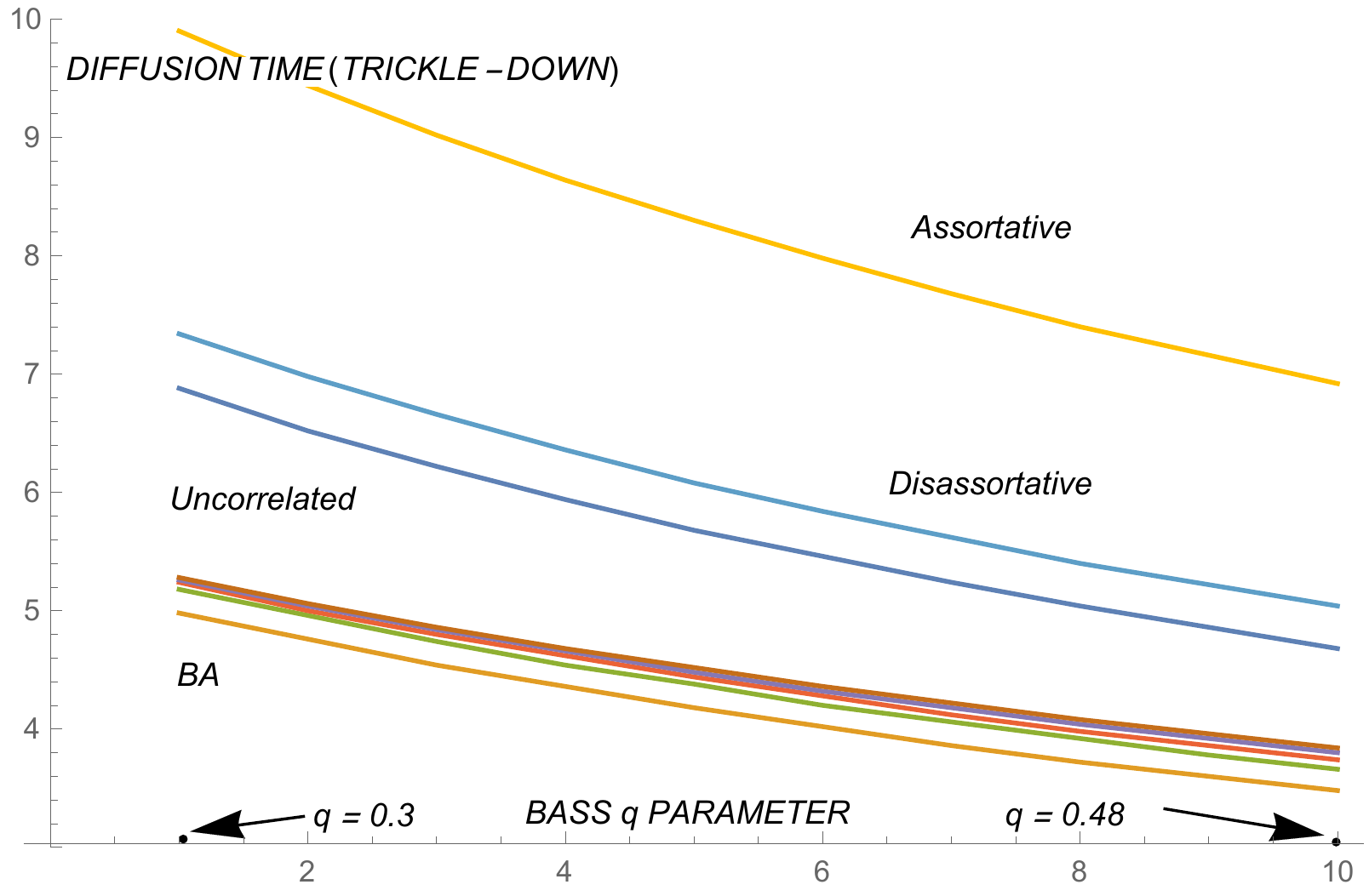}
\caption{Time of the diffusion peak (in years) for the ``trickle-down'' modified Bass model on different kinds of scale-free networks with exponent $\gamma=3$, as a function of the imitation coefficient $q$. All networks have maximum degree $n=99$. The $q$ coefficient varies in the range $0.3 - 0.48$, corresponding to a typical set of realistic values in innovation diffusion theory \cite{jiang2006virtual}. 
The publicity coefficient $p_k$ of the link class $k$, with $k=1, \ldots ,n$ varies in inverse proportion to $P(k)$ (Sect.\ \ref{heterog}).
The low-lying group of lines corresponds to BA networks with $\beta=1,2,3,4,5$ and the same assortativity coefficients as in Fig.\ \ref{fig1}. The disassortative network is built with the Newman recipe (Sect.\ \ref{matriciD}) with $d=4$ and has $r=-0.084$. The assortative network is built with our recipe (Sect.\ \ref{matriciA}), with $\alpha=1/4$, and has $r=0.827$.
} 
\label{figPVAR}
\end{center}  
\end{figure}

As can be seen also from Fig.\ \ref{figPVAR}, this has the effect of making the total diffusion faster or slower, depending on the kind of network. The differences observed in the presence of a homogeneous $p$ coefficient (Fig.\ \ref{fig1}) are now amplified. We observe that diffusion on BA networks is now always faster than on uncorrelated networks, independently from $\beta$. It is also remarkable how slow diffusion becomes on assortative networks in this case. A possible interpretation is the following: due to the low epidemic threshold of assortative networks, the targeted advertising on the hubs (which are well connected to each other) causes them to adopt very quickly, but this is not followed by easy diffusion on the whole network. The BA networks appear instead to take advantage, in the presence of advertising targeted on the hubs, both from their low threshold and from a stronger linking to the periphery.

\section{Conclusions}
\label{Conclusions}

In this work we have studied the assortativity properties of BA networks with maximum degree $n$ finite, but large (up to $n \simeq 10^4$, corresponding to a number of nodes $N \simeq 10^8$). These properties were not known in details until recently; a new decisive input has come from the exact calculation of the conditional probabilities $P(h|k)$ by Fotouhi and Rabbat \cite{fotouhi2013degree}. We have done an explicit numerical evaluation of the average nearest neighbor degree function $k_{nn}(k)$, whose behavior turns out to be peculiar and unexpected, exhibiting a coexistence of assortative and disassortative correlations, for different intervals of the node degree $k$. These results have been compared with previous estimates, both concerning the function $k_{nn}(k)$ and the Newman assortativity index $r$.

The next step has been to write the Bass innovation diffusion model on BA networks, following the mean-field scheme we have recently introduced and tested on generic scale-free networks. This allows to compute numerically the dependence of the diffusion peak time (for $n \simeq 10^2$) from the model's parameters and especially from the features of the network. We have thus compared the diffusion times on BA networks with those on uncorrelated, assortative and disassortative networks (the latter built respectively with our mathematical recipe (\ref{e15}), (\ref{Phkassort}) and with Newmans's recipe (\ref{ehkNewmandis})).

The BA networks with small values of $\beta$ ($\beta$ is the number of child nodes in the preferential attachment scheme) turn out to have the shortest diffusion time, probably due to their mixed assortative/disassortative character: diffusion appears to start quite easily among the (slightly assortative) hubs and then to proceed quickly in the (disassortative) periphery of the network. This interpretation is confirmed by the fact that in a modified ``trickle-down'' version of the model with enhanced publicity on the hubs, the anticipation effect of BA networks compared to the others is stronger and almost independent from $\beta$.

Concerning the dependence of the diffusion time on the values of the $r$ coefficient, we have found a qualitative agreement with previous results by Caldarelli et al.\ \cite{d2012robustness}.

In forthcoming work we shall analyse mathematically the construction and the properties of the mentioned families of assortative and disassortative networks, from which we only have chosen here a few samples for comparison purposes.

\section{Appendix}
\label{appendix}

We show here that the $e_{kh}$ in $(\ref{ehkNewmandis})$ satisfy $0 \le e_{kh} \le 1$ for all $h, k  = 0,...,n-1$.
If $\gamma = 2+d$ with $d>0$, 
$e_{k-1,h-1}$ can be rewritten for any $h, k  = 1,...,n$ as
\begin{equation}
e_{k-1,h-1} 
= \frac{\big( k^d + h^d \big) \sum_{j=1}^{n} {\frac{1}{j^{\gamma}}} 
-  \sum_{j=1}^{n} {\frac{1}{j^2}}}{ k^{\gamma} h^{\gamma} \, {\sum_{j=1}^{n} \frac{1}{j^2}} \, \Big({\sum_{j=1}^{n} \frac{1}{j^{\gamma}}}\Big)^2  } \, .
\label{eh-1k-1}
\end{equation}
The nonnegativity of $e_{k-1,h-1}$ is hence equivalent to that of the numerator of $(\ref{eh-1k-1})$.
This is, for all $h, k  = 1,...,n$, greater than or equal to
$$
2 \, \sum_{j=1}^{n} {\frac{1}{j^{\gamma}}}  -  \sum_{j=1}^{\infty} {\frac{1}{j^2}} \ge 2 - \frac{\pi}{6} > 0 \, .
$$
To show that $e_{k-1,h-1} \le 1$, we distinguish two cases:

i) if $k=1$ and $h=1$, the expression on the r.h.s. in $(\ref{eh-1k-1})$ is equal to
\begin{equation*}
\frac{2 \,  \sum_{j=1}^{n} {\frac{1}{j^{\gamma}}} 
-  \sum_{j=1}^{n} {\frac{1}{j^2}}}{ {\sum_{j=1}^{n} \frac{1}{j^2}} \, \Big({\sum_{j=1}^{n} \frac{1}{j^{\gamma}}}\Big)^2 } 
\le \frac{2 \,  \sum_{j=1}^{n} {\frac{1}{j^2}} 
-  \sum_{j=1}^{n} {\frac{1}{j^2}}}{ {\sum_{j=1}^{n} \frac{1}{j^2}} \, \Big({\sum_{j=1}^{n} \frac{1}{j^{\gamma}}}\Big)^2 }
\le \frac{1}{\Big({\sum_{j=1}^{n} \frac{1}{j^{\gamma}}}\Big)^2 } \le 1 \, ;
\end{equation*}

ii) otherwise, i.e. if at least one among $k$ and $h$ is greater than $1$,
the expression on the r.h.s. in $(\ref{eh-1k-1})$ is no greater than
\begin{equation*}
\frac{\big( k^d + h^d \big) \sum_{j=1}^{n} {\frac{1}{j^{2+d}}} }{ k^{2+d} h^{2+d} \, {\sum_{j=1}^{n} \frac{1}{j^2}} \, \Big({\sum_{j=1}^{n} \frac{1}{j^{2+d}}}\Big)^2  } 
\le \frac{\big( k^d + h^d \big)}{k^d h^d} \, \frac{1}{ k^2 h^2 } \le 1 \, .
\end{equation*}

\bibliographystyle{unsrt}
\bibliography{jb3refs}
 
\end{document}